\documentclass[10pt,showpacs,amsmath,amssymb]{article}
\usepackage{amsfonts}
\usepackage{mathdots}
\usepackage{color}

\begin{document}

\title{Superforms and the $\mathbb{C}P^{N-1}$ supersymmetric sigma model}

\author{Laurent Delisle}

\author{Laurent Delisle${}^{1,2}$}

\footnotetext[1]{Institut de math\'ematiques de Jussieu-Paris Rive Gauche, UP7D-Campus des Grands Moulins, B\^atiment Sophie Germain, Cases 7012, 75205 Paris Cedex 13.}
\footnotetext[2]{email:laurent.delisle@imj-prg.fr}

\date{\today}

\maketitle

\begin{abstract}
We present a characterisation of Maurer-Cartan $1$-superforms associated to the two-dimensional supersymmetric $\mathbb{C}P^{N-1}$ sigma model. We, then, solve the associated linear spectral problem and use its solutions to describe an integrable system for a $su(N)$-valued map.
\end{abstract}

%%%%%%%%%%%%%%%section 1%%%%%%%%%%%%%

\section{Introduction}

The study of integrable systems is the subject of great interest in the mathematics and physics communities. In particular, the solutions of the bosonic and supersymmetric two-dimensional $\mathbb{C}P^{N-1}$ sigma models have been considered and used to investigate the geometry of surfaces immersed in the Lie algebra $su(N)$ \cite{delisle,delisle1,hussin,delisle2,delisle3,delisle4,grundland,bolton,jiao,peng,peng1,goldstein,grundland1,zakrzewski}. Indeed, the present author has considered these surfaces of constant Gaussian curvature and gave a classification of solutions leading to such surfaces in terms of the Veronese curve \cite{delisle1,delisle2,delisle3,delisle4}.

In the bosonic model, invariant recurrence relations linking the solutions of the associated linear spectral problem, the solutions of the model and the surfaces immersed in the Lie algebra $su(N)$ were given \cite{goldstein}. These links were established using the Sym-Tafel \cite{goldstein,grundland1} and Weierstrass immersion \cite{kono} formulas. In the supersymmetric model, such an analysis is absent and this was, until now, partially due to the absence of a systematic procedure of constructing non-holomorphic solutions. This problem was recently solved in \cite{delisle}, where the authors generalized the bosonic holomorphic method. Using these new solutions, we were interested to study, in the supersymmetric context, the solutions of the associated linear spectral problem, the surfaces immersed in the Lie algebra $su(N)$ and the links between them. The starting point to this analysis is based on the following result on superharmonic maps into symmetric spaces \cite{khemar,odea} (noting that $\mathbb{C}P^{N-1}$ is indeed symmetric):
%The study of integrability conditions of mathematics and physics models has been subject of great interest. In particular, the solutions of the two-dimensional supersymmetric $\mathbb{C}P^{N-1}$ sigma model have been recently considered  by the author. In the bosonic model, the solutions were used to study, as for example, the geometry of surfaces immersed in the Lie algebra $su(N)$ and the solutions of the associated linear spectral problem. Indeed, in \cite{goldstein}, the authors have obtained different recurrence relations linking those quantities. We propose, in this paper, to discuss these subjects for the supersymmetric extension of the $\mathbb{C}P^{N-1}$ model from the supergeometry point of view. Our approach is based on the characterisation of Maurer-Cartan $1$-superforms in the study of superharmonic maps into symmetric spaces:

\noindent\textbf{Theorem} (Khemar) \cite{khemar} Let $\alpha$ be a $1$-superform on $\mathbb{C}^{1\vert 1}$ with values in the Lie algebra $\mathfrak{g}$ of the Lie group $G$. Then there exists $\mathcal{F}\,:\,\mathbb{C}^{1\vert 1}\longrightarrow G$ such that $\alpha=\mathcal{F}^{-1}d\mathcal{F}$ if and only if $d\alpha+\alpha\wedge\alpha=0$.

This theorem describes the Maurer-Cartan $1$-superform and, in this paper, we propose to characterize these superforms for the supersymmetric $\mathbb{C}P^{N-1}$ sigma model. This characterisation is important for the study of the integrability of the model. These superforms can then be used to determine a linear spectral problem \cite{khemar} and the solutions of this problem, associated to the new solutions of the model, is of great interest in determining the Weierstrass immersion formula from the Sym-Tafel formula \cite{goldstein,grundland1}. These formulas will give $su(N)$-valued maps (or surfaces) which solves an integrable system of first order equations. Note that these surfaces were extensively studied in the bosonic model \cite{delisle2,delisle3,delisle4,grundland,bolton,jiao,peng,peng1,goldstein,grundland1,zakrzewski}.

%We propose, in this paper, to describe the Maurer-Cartan $1$-superforms related to the supersymmetric $\mathbb{C}P^{N-1}$ sigma model and to construct, explicitly, the map $\mathcal{F}$. This characterisation will be used to study the associated linear spectral problem and its link with a system of linear equations for a $su(N)$-valued quantity $\mathbf{X}$.
The paper is divided as follows. In section 2, we recall, in order for the paper to be self-contained, the definition of the two-dimensional supersymmetric $\mathbb{C}P^{N-1}$ sigma model, its gauge-invariant representation and, briefly, present the new technique of solving the Euler-Lagrange equations associated to this model. Using those newly discovered solutions, we give nice representations of the Lagrangian and topological densities in terms of the supersymmetric $\mathbb{C}^{1\vert 1}$ Laplacian operator \cite{zakrzewski}. The following section gives a description of superforms in $\mathbb{C}^{1\vert 1}$ and the conditions for which they are closed, exact and $su(N)$-valued. Section 4 is devoted to the complete characterisation of the Maurer-Cartan $1$-superform, the explicit solutions of the associated linear spectral problem and the construction of surfaces using these solutions. Finally, we conclude with some future outlooks.
%In section 2, we recall the definition of the supersymmetric $\mathbb{C}P^{N-1}$ sigma model, its gauge-invariant representation and present the recently developed technique of solving the Euler-Lagrange equations associated to this model. The following section is devoted to superforms in $\mathbb{C}^{1\vert 1}$ and the conditions for which they are closed, exact or/and $su(N)$-valued. Section 4 is devoted to the characterization of Maurer-Cartan superforms and the linear spectral problem associated to the model. Finally, we conclude with some future outlooks.
\section{The supersymmetric $\mathbb{C}P^{N-1}$ sigma model}

In order for the paper to be self-contained, we recall the definition of the two-dimensional supersymmetric $\mathbb{C}P^{N-1}$ sigma model and its gauge-invariant representation \cite{zakrzewski}. We, also, describe a special class of solutions for this model \cite{delisle} and use them to deduce, in subsequent sections, different properties.

\subsection{The model}

The two-dimensional supersymmetric $\mathbb{C}P^{N-1}$ sigma model \cite{zakrzewski} describes the collection of bosonic superfields $\Phi$ defined on the two-dimensional Euclidean complex superspace $\mathbb{C}^{1\vert 1}$ \cite{dewitt,cornwell} of local coordinates $(x_+,x_-,\theta_+,\theta_-)$ with values in the Grassmannian manifold $\mathbb{C}P^{N-1}$. In this superspace formulation, $(x_+,x_-)$ are local coordinates of the complex plane $\mathbb{C}$  with $x_+^{\dagger}=x_-$ and $(\theta_+,\theta_-)$ are complex odd Grassmann variables with $\theta_+^{\dagger}=\theta_-$ satisfying the anti-commutation relations:
\begin{equation}
\theta_+\theta_-+\theta_-\theta_+=0,\quad \theta_+^2=\theta_-^2=0.
\end{equation} 
The superfields $\Phi$ are described by $N$-components vectors (\textit{i.e.} $\Phi\in (\mathbb{C}^{1\vert 1})^N$) satisfying the non-linear constraint
\begin{equation}
\vert\Phi\vert^2=\Phi^{\dagger}\Phi=1.
\label{constraintCPN}
\end{equation}
The classical solutions $\Phi$ of the model are the critical points of finite energy of the action functional $S(\Phi)=\int d^2\theta d^2x\mathcal{L}(\Phi)$ where the Lagrangian density is given by
\begin{equation}
\mathcal{L}(\Phi)=2(\vert\check{D}_+\Phi\vert^2-\vert\check{D}_-\Phi\vert^2).
\label{lagrangian}
\end{equation}
The covariant derivatives $\check{D}_{\pm}$ are constructed from the gauge invariance of the action $S(\Phi)$ under the transformation $\Phi^{\prime}=V\Phi U$ where $U\in U(1)$ and $V\in U(N)$ are, respectively, local and global gauge transformations of the unitary group. They are explicitly given as
\begin{equation}
\check{D}_{\pm}\Lambda=\check{\partial}_{\pm}\Lambda-\Lambda(\Phi^{\dagger}\check{\partial}_{\pm}\Phi),
\end{equation}
where the operators $\check{\partial}_{\pm}$ are the supersymmetric-invariant vector fields defined by
\begin{equation}
\check{\partial}_{\pm}=-i\partial_{\theta_{\pm}}+\theta_{\pm}\partial_{\pm}
\label{superder}
\end{equation}
satisfying $\check{\partial}_{\pm}^2=-i\partial_{x_{\pm}}\equiv -i\partial_{\pm}$. Using the least action principle, we deduce the Euler-Lagrange equations given by
\begin{equation}
\check{D}_+\check{D}_-\Phi+\vert\check{D}_-\Phi\vert^2\Phi=0,\quad \vert\Phi\vert^2=1
\label{EL}
\end{equation}
and, in order to get finite action solutions $\Phi$, we impose the boundary conditions
\begin{equation}
\check{D}_{\pm}\Phi\longrightarrow 0,\quad \hbox{as}\quad x_+x_-\longrightarrow \infty.
\end{equation}
An other important object, which will be discussed in more details in section 2.4, associated to the model is the topological density $\mathcal{Q}$ defined as
\begin{equation}
\mathcal{Q}(\Phi)=2(\vert \check{D}_+\Phi\vert^2+\vert\check{D}_-\Phi\vert^2).
\label{topological}
\end{equation}

\subsection{The gauge-invariant representation}

In this section, we recall the gauge-invariant formulation of the supersymmetric $\mathbb{C}P^{N-1}$ sigma model in terms of hermitian and orthogonal projectors $\mathbb{P}$ of rank one \cite{zakrzewski}. These projectors satisfy the properties
\begin{equation}
\mathbb{P}^2=\mathbb{P}^{\dagger}=\mathbb{P},\quad \hbox{Tr}(\mathbb{P})=1
\label{orthoproj}
\end{equation}
and such projectors may be constructed from a solution $\Phi$ of the model as
\begin{equation}
\mathbb{P}=\Phi\otimes \Phi^{\dagger}.
\end{equation}
Indeed, the properties (\ref{orthoproj}) are satisfied following the non-linear constraint (\ref{constraintCPN}). In this representation, the Euler-Lagrange equations (\ref{EL}) are equivalently written as
\begin{equation}
[\check{\partial}_+\check{\partial}_-\mathbb{P},\mathbb{P}]=0,\quad \mathbb{P}^2=\mathbb{P},
\label{ELP}
\end{equation}
where $[\cdot,\cdot]$ is the usual matrix commutator. We notice that this formalism is gauge-invariant since, under gauge transformation $\Phi^{\prime}=V\Phi U$, the projector $\mathbb{P}$ changes as $\mathbb{P}^{\prime}=V\mathbb{P}V^{\dagger}$ which is free from the local gauge unitary transformation $U$. An important observation, which will become handy, is that the Euler-Lagrange equations may be reformulated as a super-conservation law \cite{delisle,delisle1,hussin} as
\begin{equation}
\check{\partial}_-\varphi_+-\check{\partial}_+\varphi_+^{\dagger}=0,\quad \varphi_+=i[\mathbb{P},\check{\partial}_+\mathbb{P}].
\label{conserve}
\end{equation}

\subsection{A special class of solutions}

In this section, we recall the construction, based on the holomorphic method, of a special class of solutions of the Euler-Lagrange equations \cite{delisle}. 

Consider a set of linearly independent holomorphic (\textit{i.e.} functions of $(x_+,\theta_+)$) $N$-components vectors $\{\psi_0,\psi_1,\cdots,\psi_{N-1}\}$ defined by the system of equations
\begin{equation}
\epsilon_1\psi_1=\check{\partial}_+\psi_0,\quad \epsilon_2\psi_2=\check{\partial}_+\psi_1,\quad \cdots,\quad \epsilon_{N-1}\psi_{N-1}=\check{\partial}_+\psi_{N-2},
\end{equation}
where $\epsilon_j=\epsilon_j(x_+,\theta_+)$ are scalar fermionic functions for $j=1,2,\cdots,N-1$. From the Gram-Schmidt procedure, we construct, from this set, a new set of orthogonal vectors $\{z_0,z_1,\cdots,z_{N-1}\}$ defined as
\begin{equation}
z_0=\psi_0,\quad z_j=\left(\mathbb{I}-\sum_{k=0}^{j-1}\mathbb{P}_k\right)\psi_j,\quad j=1,2,\cdots, N-1,
\end{equation}
where the quantities $\mathbb{P}_k$ are rank one, hermitian and orthogonal projectors defined as
\begin{equation}
\mathbb{P}_k=\Phi_k\otimes \Phi_k^{\dagger}=\frac{z_k\otimes z_k^{\dagger}}{\vert z_k\vert^2},\quad \Phi_k=\frac{z_k}{\vert z_k\vert},
\label{solutionmodel}
\end{equation}
for $k=0,1,\cdots,N-1$. As shown in \cite{delisle}, $\Phi_k$ and $\mathbb{P}_k$ are solutions, respectively, of the Euler-Lagrange equations (\ref{EL}) and (\ref{ELP}) for all $k$. The solutions corresponding to $k=0$ and $k=N-1$ are known, respectively, as holomorphic and anti-holomorphic while the others are called non-holomorphic or mixed. Furthermore, the vectors $z_k$ satisfy the convenient properties
\begin{equation}
\check{\partial}_+\left(\frac{z_j}{\vert z_j\vert^2}\right)=\epsilon_{j+1}\frac{z_{j+1}}{\vert z_j\vert^2},\quad \check{\partial}_-z_j=-\epsilon_j^{\dagger}\frac{\vert z_j\vert^2}{\vert z_{j-1}\vert^2}z_{j-1}
\label{propz}
\end{equation}
for $j=0,1,\cdots,N-1$.

\subsection{The Lagrangian and topological densities}
In this section, we give representations of the topological and Lagrangian densities associated to these recently discovered solutions in terms of the supersymmetric $\mathbb{C}^{1\vert 1}$ Laplacian operator. Indeed, given a solution $\Phi_k$ as defined in (\ref{solutionmodel}), we have that the covariant derivatives $\check{D}_{\pm}$ satisfy 
\begin{equation}
\vert \check{D}_{\pm}\Phi_k\vert^2=\pm\vert\epsilon_{k+\frac12\pm\frac12}\vert^2\frac{\vert z_{k+\frac12\pm \frac12}\vert^2}{\vert z_{k-\frac{1}{2}\pm\frac12}\vert^2}
\end{equation}
from which we deduce the explicit expressions of the Lagrangian (\ref{lagrangian}) and topological (\ref{topological}) densities as
\begin{eqnarray}
\mathcal{L}_k&=&\mathcal{L}(\Phi_k)=2\left(\vert\epsilon_{k+1}\vert^2\frac{\vert z_{k+1}\vert^2}{\vert z_k\vert^2}+\vert\epsilon_k\vert^2\frac{\vert z_k\vert^2}{\vert z_{k-1}\vert^2}\right),\\
\mathcal{Q}_k&=&\mathcal{Q}(\Phi_k)=2\left(\vert\epsilon_{k+1}\vert^2\frac{\vert z_{k+1}\vert^2}{\vert z_k\vert^2}-\vert\epsilon_k\vert^2\frac{\vert z_k\vert^2}{\vert z_{k-1}\vert^2}\right).
\end{eqnarray}
As in the bosonic model \cite{delisle3} and following the properties (\ref{propz}), we have that
\begin{equation}
\check{\partial}_-\check{\partial}_+\log\vert z_k\vert^2=\vert \epsilon_{k+1}\vert^2\frac{\vert z_{k+1}\vert^2}{\vert z_k\vert^2}-\vert \epsilon_k\vert^2\frac{\vert z_k\vert^2}{\vert z_{k-1}\vert^2}=\frac{1}{2}\mathcal{Q}_k
\end{equation}
and, for the Lagrangian density $\mathcal{L}_k$, we easily deduce that
\begin{equation}
\mathcal{L}_k=2\check{\partial}_-\check{\partial}_+\log\left(\vert z_k\vert^2\vert z_{k-1}\vert^4\vert z_{k-2}\vert^4\cdots\vert z_0\vert^4\right).
\end{equation}
We can go further and enumerate some nice properties of the quantities $\log\vert z_k\vert^2$ or, equivalently, of the topological densities $\mathcal{Q}_k$. Indeed, we can make the observations:
\begin{enumerate}
\item The first property that one can deduce about the topological densities $\mathcal{Q}_k$ is that they sum to zero, \textit{i.e.}
\begin{equation}
\sum_{k=0}^{N-1}\mathcal{Q}_k=0\quad \iff\quad \sum_{k=0}^{N-1}\check{\partial}_-\check{\partial}_+\log\vert z_k\vert^2=0.
\label{sumzero}
\end{equation}
Using a Taylor expansion, we may thus write
\begin{equation}
\sum_{k=0}^{N-1}\log\vert z_k\vert^2=A_0(x_+,x_-)+i\theta_+\zeta_+(x_+,x_-)+i\theta_-\zeta_-(x_+,x_-)-\theta_+\theta_-A_{+-}(x_+,x_-),
\end{equation}
where $A_0$, $A_{+-}$ are bosonic scalar functions and $\zeta_+$, $\zeta_-$ are fermionic scalar functions. From (\ref{sumzero}), we deduce that
\begin{equation}
A_0(x_+,x_-)=a_0(x_+)+a_1(x_-),\quad \zeta_+=\zeta_+(x_+),\quad \zeta_-=\zeta_-(x_-),\quad A_{+-}=0.
\end{equation}
Using the properties (\ref{propz}), we obtain
\begin{equation}
\check{\partial}_+\log\vert z_k\vert^2=\epsilon_{k+1}\left(\frac{z_k}{\vert z_k\vert^2}\right)^{\dagger}\psi_{k+1}-\epsilon_k\left(\frac{z_{k-1}}{\vert z_{k-1}\vert^2}\right)^{\dagger}\psi_{k},
\end{equation}
and this allows us to deduce that
\begin{equation}
\sum_{k=0}^{N-1}\check{\partial}_+\log\vert z_k\vert^2=0=\zeta_+(x_+)+\theta_+\partial_+a_0(x_+)
\end{equation}
which implies that $\zeta_+=0$ and $a_0$ is a constant and, by symmetry, we also have $\zeta_-=0$ and $a_1$ is a constant. We thus obtain the following property of the functions $z_k$:
\begin{equation}
\sum_{k=0}^{N-1}\log\vert z_k\vert^2=\kappa_{N-1},
\end{equation}
where $\kappa_{N-1}$ is a constant.
\item The second observation that one can make is that the topological density $\mathcal{Q}_k$ may be represented as the difference of two quantities \cite{din}. Indeed, defining
\begin{equation}
\xi_+^k=\sum_{i=0}^{k-1}(\mathcal{Q}_i-\mathcal{Q}_k),\quad \xi_-^{k}=\sum_{i=k+1}^{N-1}(\mathcal{Q}_k-\mathcal{Q}_i),
\end{equation}
we can show, following the properties found in point 1, that
\begin{equation}
\xi_+^k-\xi_-^k=-N\mathcal{Q}_k.
\label{atiyah}
\end{equation}
This follows from the definition of the quantities $\xi_+^k$ and $\xi_-^{k}$, where we deduce that
\begin{equation}
\xi_+^k-\xi_-^k=(1-N)\mathcal{Q}_k+2\sum_{i\neq k}\check{\partial}_-\check{\partial}_+\log\vert z_i\vert^2=(1-N)\mathcal{Q}_k+\sum_{i\neq k}\mathcal{Q}_i
\end{equation}
and, from the property (\ref{sumzero}), this implies the relation (\ref{atiyah}).
\end{enumerate}

\section{Exact, closed and $su(N)$-valued superforms}

In this section, we describe differential forms in superspace \cite{dewitt,cornwell}, called superforms, and the conditions for which these superforms are exact, closed and $su(N)$-valued.

Let $f=f(x_+,x_-,\theta_+,\theta_-)$ be a function of superspace or, equivalently, a $0$-superform, we define the exterior derivative $d$ by its action on $f$ as
\begin{equation}
df=dx_+\partial_+f+dx_-\partial_-f+d\theta_+\partial_{\theta_+}f+d\theta_-\partial_{\theta_-}f.
\end{equation}
Using the superderivatives (\ref{superder}) as $\partial_{\theta_{\pm}}=i(\check{\partial}_{\pm}-\theta_{\pm}\partial_{\pm})$, we can re-write this expression as
\begin{equation}
df=A_+\partial_+f+A_-\partial_-f+\Pi_+\check{\partial}_+f+\Pi_-\check{\partial}_-f,
\end{equation}
where $A_{\pm}=dx_{\pm}-i(d\theta_{\pm})\theta_{\pm}$ and $\Pi_{\pm}=id\theta_{\pm}$ are, respectively, bosonic and fermionic $1$-superforms. The exterior derivative satisfies $d^2=0$ and the following properties:
\begin{eqnarray}
dx_{\mu}\wedge dx_{\nu}&=&-dx_{\nu}\wedge dx_{\mu},\quad x_{\mu} dx_{\nu}=(dx_{\nu})x_{\mu},\quad d(x_{\mu}dx_{\nu})=dx_{\mu}\wedge dx_{\nu},\nonumber\\
d\theta_{\mu}\wedge dx_{\nu}&=&-dx_{\nu}\wedge d\theta_{\mu},\quad\theta_{\mu}dx_{\nu}=(dx_{\nu})\theta_{\mu},\quad d(\theta_{\mu}dx_{\nu})=d\theta_{\mu}\wedge dx_{\nu},\\
d\theta_{\mu}\wedge d\theta_{\nu}&=&d\theta_{\nu}\wedge d\theta_{\mu},\quad \theta_{\mu}d\theta_{\nu}=-(d\theta_{\nu})\theta_{\mu},\quad d(\theta_{\mu}d\theta_{\nu})=d\theta_{\mu}\wedge d\theta_{\nu},\nonumber
\end{eqnarray}
for all $\mu,\nu=+,-$ and, unlike bosonic models, we have that $d\theta_{\mu}\wedge d\theta_{\mu}\neq0$. From these properties, we may define the action of the exterior derivative $d$ on $1$-superforms. Concerning this matter, a general $1$-superform $\alpha$ is given as
\begin{equation}
\alpha=A_+a_++A_-a_-+\Pi_+\pi_++\Pi_-\pi_-,
\end{equation}
where $a_{\pm}$ and $\pi_{\pm}$ are, respectively, bosonic and fermionic functions of $\mathbb{C}^{1\vert 1}$. For the purpose of the paper, we enumerate the conditions for which a $1$-superform is $su(N)$-valued, closed and exact. Before doing so, let us make some preliminary calculations. We have
\begin{eqnarray}
d\alpha&=&-\Pi_+\wedge\Pi_+(ia_++\check{\partial}_+\pi_+)-\Pi_-\wedge\Pi_-(ia_-+\check{\partial}_-\pi_-)-\Pi_+\wedge\Pi_-(\check{\partial}_+\pi_-+\check{\partial}_-\pi_+)\nonumber\\
&+&A_+\wedge A_-(\partial_+a_--\partial_-a_+)-A_+\wedge\Pi_+(\check{\partial}_+a_+-\partial_+\pi_+)-A_+\wedge\Pi_-(\check{\partial}_-a_+-\partial_+\pi_-)\nonumber\\
&-&A_-\wedge\Pi_-(\check{\partial}_-a_--\partial_-\pi_-)-A_-\wedge \Pi_+(\check{\partial}_+a_--\partial_-\pi_+)
\end{eqnarray}
and
\begin{eqnarray}
\alpha\wedge\alpha &=&-A_+\wedge\Pi_+[\pi_+,a_+]-A_+\wedge\Pi_-[\pi_-,a_+]-A_-\wedge\Pi_-[\pi_-,a_-]-A_-\wedge\Pi_+[\pi_+,a_-]\nonumber\\
&-&\Pi_+\wedge\Pi_+ \pi_+^2-\Pi_-\wedge\Pi_- \pi_-^2-\Pi_+\wedge\Pi_-\{\pi_+,\pi_-\}+A_+\wedge A_- [a_+,a_-].
\end{eqnarray}
The last expression, present in the theorem of Khemar \cite{khemar}, will be used in the subsequent sections. The first expression is useful in proving the following \cite{dewitt,cornwell}:
\begin{enumerate}
\item $\alpha\in su(N)$: The $1$-superform $\alpha$ is $su(N)$-valued if it is defined by a $N\times N$ matrix satisfying $\alpha^{\dagger}=-\alpha$ and $\hbox{Tr}(\alpha)=0$. Following the facts that $A_+^{\dagger}=A_-$ and $\Pi_+^{\dagger}=-\Pi_-$, we have
\begin{equation}
\alpha^{\dagger}=A_+a_-^{\dagger}+A_-a_+^{\dagger}+\Pi_+\pi_-^{\dagger}+\Pi_-\pi_+^{\dagger}=-\alpha,\quad \hbox{Tr}(\alpha)=0
\end{equation}
if and only if $a_+^{\dagger}=-a_-$ and $\pi_+^{\dagger}=-\pi_-$ together with $\hbox{Tr}(a_{\pm})=\hbox{Tr}(\pi_{\pm})=0$. As a conclusion, a general $1$-superform $\alpha$ is $su(N)$-valued if it as the following form
\begin{equation}
\alpha=A_+a_+-A_-a_+^{\dagger}+\Pi_+\pi_+-\Pi_-\pi_+^{\dagger}
\label{suNform}
\end{equation}
together with $\hbox{Tr}(a_{\pm})=\hbox{Tr}(\pi_{\pm})=0$.
\item Closed superforms: A $k$-superform $\beta$ is closed if it satisfies $d\beta=0$. In the case of the $1$-superform $\alpha\in su(N)$, we have that $\alpha$ is closed if
\begin{equation}
a_+=i\check{\partial}_+\pi_+,\quad \check{\partial}_-\pi_+-\check{\partial}_+\pi_+^{\dagger}=0.\label{closedprop}
\end{equation}
\item Exact superforms: A $k$-superform $\beta$ is exact if there exist a $(k-1)$-superform $\gamma$ such that $\beta=d\gamma$. In the case of the $1$-superform $\alpha\in su(N)$, we have that $\alpha$ is exact if there exist a $0$-superform $f$ such that $\alpha=df$. This equation is equivalent to
\begin{equation}
\partial_+f=a_+,\quad \partial_-f=-a_+^{\dagger},\quad \check{\partial}_+f=\pi_+,\quad \check{\partial}_-f=-\pi_+^{\dagger}.
\end{equation}
This system is compatible if and only if the conditions given in (\ref{closedprop}) are satisfied. This is compatible with the fact that an exact superform is automatically closed.
\end{enumerate}

\section{Maurer-Cartan superforms in superspace $\mathbb{C}^{1\vert 1}$}

In this section, following the theorem of Khemar \cite{khemar}, we describe the $1$-superforms in superspace $\mathbb{C}^{1\vert 1}$ associated to the supersymmetric $\mathbb{C}P^{N-1}$ model which are of the Maurer-Cartan form. In the formulation of the theorem of Khemar and from the results of the bosonic model \cite{grundland}, we choose $\mathfrak{g}=su(N)$ and $G=SU(N)\subset U(N)$. In this case, the $1$-superforms $\alpha$ are completely described as in (\ref{suNform}) and are solutions of the Maurer-Cartan equation
\begin{equation}
d\alpha+\alpha\wedge \alpha=0
\end{equation}
if and only if the following system of equations is satisfied
\begin{eqnarray}
ia_++\check{\partial}_+\pi_++\pi_+^2&=&0,\label{eq1}\\
\check{\partial}_-\pi_+-\check{\partial}_+\pi_+^{\dagger}-\{\pi_+,\pi_+^{\dagger}\}&=&0,\label{eq2}\\
\partial_+a_+^{\dagger}+\partial_-a_++[a_+,a_+^{\dagger}]&=&0,\label{eq3}\\
\partial_+\pi_+-\check{\partial}_+a_++[a_+,\pi_+]&=&0,\label{eq4}\\
\partial_+\pi_+^{\dagger}+\check{\partial}_-a_++[a_+,\pi_+^{\dagger}]&=&0.\label{eq5}
\end{eqnarray}
The first equation (\ref{eq1}) of this system gives an explicit expression of the bosonic quantity $a_+$ in terms of the fermionic quantity $\pi_+$. Indeed, we get
\begin{equation}
a_+=i(\check{\partial}_+\pi_++\pi_+^2).
\label{aplus}
\end{equation}  
Let us now consider equation (\ref{eq2}) of the above system and seek for a solution of the form
\begin{equation}
\pi_+=i\gamma_+\varphi_+,
\label{piplus}
\end{equation}
where $\gamma_+$ is a bosonic scalar and $\varphi_+$ was defined in equation (\ref{conserve}). For $\mathbb{P}$ a solution of the Euler-Lagrange equations (\ref{ELP}), direct calculations lead to
\begin{equation}
\check{\partial}_-\pi_+=\gamma_+([\check{\partial}_-\check{\partial}_+\mathbb{P},\mathbb{P}]-\{\check{\partial}_+\mathbb{P},\check{\partial}_-\mathbb{P}\})=-\gamma_+\{\check{\partial}_+\mathbb{P},\check{\partial}_-\mathbb{P}\}
\end{equation}
and
\begin{equation}
\pi_+\pi_+^{\dagger}=\gamma_+\gamma_+^{\dagger}\check{\partial}_+\mathbb{P}\check{\partial}_-\mathbb{P}
\end{equation}
from which the equation (\ref{eq2}) of the system becomes
\begin{equation}
(\gamma_++\gamma_+^{\dagger}+\gamma_+\gamma_+^{\dagger})\{\check{\partial}_+\mathbb{P},\check{\partial}_-\mathbb{P}\}=0\quad \iff\quad \gamma_++\gamma_+^{\dagger}+\gamma_+\gamma_+^{\dagger}=0.
\end{equation}
This equation may be solved in terms of a free complex parameter $\lambda$. Indeed, we obtain
\begin{equation}
\gamma_+=\lambda-1,\quad \vert\lambda\vert^2=1.
\end{equation}
We may thus calculate the explicit expressions of $a_+$ and $\pi_+$ given, respectively, in (\ref{aplus}) and (\ref{piplus}). We get
\begin{equation}
a_+^{\lambda}=(\lambda-1)[\partial_+\mathbb{P},\mathbb{P}]-i(\lambda^2-1)(\check{\partial}_+\mathbb{P})^2,\quad \pi_+^{\lambda}=(\lambda-1)[\check{\partial}_+\mathbb{P},\mathbb{P}],
\label{appiplambda}
\end{equation}
where we have added the superscript $\lambda$ to emphasise the dependence on this free complex parameter. It remains to show that equations (\ref{eq3}), (\ref{eq4}) and (\ref{eq5}) are satisfied.

\subsection{The special case of $\lambda=-1$}

In this special case, we have that the expressions given in (\ref{appiplambda}) reduce to
\begin{equation}
a_+^{-1}=-2[\partial_+\mathbb{P},\mathbb{P}],\quad \pi_+^{-1}=-2[\check{\partial}_+\mathbb{P},\mathbb{P}]
\end{equation}
and, it is direct calculations, to show that the system of equations (\ref{eq1})-(\ref{eq5}) are satisfied. Furthermore, with such expressions, the $1$-superform $\alpha$ may be compactly written as
\begin{equation}
\alpha=-2[d\mathbb{P},\mathbb{P}]
\label{Maurerform}
\end{equation}
and, from the theorem of Khemar \cite{khemar}, this $1$-superform is of the Maurer-Cartan form. This means that there exist a map $\mathcal{F}\,:\,\mathbb{C}^{1\vert 1}\longrightarrow G\subset U(N)$ such that $\alpha$ is given as
\begin{equation}
\alpha=\mathcal{F}^{-1}d\mathcal{F}.
\end{equation}
The map $\mathcal{F}$ can be explicitly constructed as
\begin{equation}
\mathcal{F}=Q(2\mathbb{P}-\mathbb{I}),
\label{Flambda1}
\end{equation}
where $Q\in U(N)$ is a constant matrix. Indeed, we have from the properties of the projector $\mathbb{P}$, given in (\ref{orthoproj}), that
\begin{equation}
\mathcal{F}^{\dagger}\mathcal{F}=4\mathbb{P}^2-4\mathbb{P}+\mathbb{I}=\mathbb{I},
\end{equation}
which is equivalent to say that $\mathcal{F}\in U(N)$ and $\mathcal{F}^{-1}=(2\mathbb{P}-\mathbb{I})Q^{\dagger}$. Using the orthogonality property of $\mathbb{P}$, we have that $d\mathbb{P}=d\mathbb{P}^2=(d\mathbb{P})\mathbb{P}+\mathbb{P}d\mathbb{P}$ from which we deduce that
\begin{equation}
\mathcal{F}^{-1}d\mathcal{F}=2(2\mathbb{P}-\mathbb{I})d\mathbb{P}=2(2\mathbb{P}d\mathbb{P}-d\mathbb{P})=2[\mathbb{P},d\mathbb{P}]=\alpha.
\end{equation}
As a final comment, in this section, we may choose $Q$ in (\ref{Flambda1}) so that $\mathcal{F}\in G=SU(N)$. Indeed, this follows directly from the fact that $\det(2\mathbb{P}-\mathbb{I})=(-1)^{N-1}$.

\subsection{The general case}
In this section, we extend our analysis of the previous section without fixing $\lambda=-1$. Indeed, we showed that $\pi_{+}^{\lambda}$, given in (\ref{appiplambda}), solves equation (\ref{eq2}). From this fact, let us consider the system of first order equations given as
\begin{equation}
\check{\partial}_+\mathcal{F}_{\lambda}=\mathcal{F}_{\lambda}\pi_+^{\lambda},\quad \check{\partial}_-\mathcal{F}_{\lambda}=-\mathcal{F}_{\lambda}(\pi_+^{\lambda})^{\dagger},
\label{speceq}
\end{equation}
where $\mathcal{F}_{\lambda}\,:\,\mathbb{C}^{1\vert 1}\longrightarrow U(N)$. We can observe that the compatibility of the set of equations (\ref{speceq}),  $\{\check{\partial}_+,\check{\partial}_-\}\mathcal{F}_{\lambda}=0$, is equivalent to equation (\ref{eq2}). In order to solve this system, we use the known solutions of the model $\mathbb{P}_k$, given in (\ref{solutionmodel}), to construct a $U(N)$-valued map $\mathcal{F}_{\lambda,k}$. For the first equation of system (\ref{speceq}), we seek a solution $\mathcal{F}_{\lambda,k}$ of the equation 
\begin{equation}
\mathcal{F}_{\lambda,k}^{-1}\check{\partial}_+\mathcal{F}_{\lambda,k}=\pi_+^{\lambda,k}=(\lambda-1)[\check{\partial}_+\mathbb{P}_k,\mathbb{P}_k],\quad \forall\lambda\in S^1.
\label{tosolve}
\end{equation}
To solve this equation, we make the following ansatz
\begin{equation}
\mathcal{F}_{\lambda,k}=Q_k\left(\mathbb{I}+\sum_{j=0}^k\beta_j(\lambda)\mathbb{P}_j\right),
\end{equation}
where $Q_k\in U(N)$ is a constant matrix and the functions $\beta_j=\beta_j(\lambda)$ are to be determined for all $j=0,1,\cdots,k$. For $\beta_j\neq-1$, it is direct to show that the inverse of the matrix $\mathcal{F}_{\lambda,k}$ is given by
\begin{equation}
\mathcal{F}_{\lambda,k}^{-1}=\left(\mathbb{I}-\sum_{j=0}^k\left(\frac{\beta_j(\lambda)}{\beta_j(\lambda)+1}\right)\mathbb{P}_j\right)Q_k^{\dagger}.
\end{equation}
Note here that in the case where $\beta_j=-1$ for all $j$, then $\mathcal{F}_{\lambda,k}\notin U(N)$. So, this case have to be rejected. Preliminary calculations lead to the expression
\begin{equation}
\check{\partial}_+\mathcal{F}_{\lambda,k}=Q_k\left(\beta_k\epsilon_{k+1}\frac{z_{k+1}\otimes z_k^{\dagger}}{\vert z_k\vert^2}+\sum_{m=1}^{k}\epsilon_m(\beta_{m-1}-\beta_m)\frac{z_m\otimes z_{m-1}^{\dagger}}{\vert z_{m-1}\vert^2}\right),
\end{equation}
from which we deduce the left-hand side of equation (\ref{tosolve}):
\begin{equation}
\mathcal{F}_{\lambda,k}^{-1}\check{\partial}_+\mathcal{F}_{\lambda,k}=\beta_k\epsilon_{k+1}\frac{z_{k+1}\otimes z_k^{\dagger}}{\vert z_k\vert^2}+\sum_{m=1}^k\frac{\epsilon_m(\beta_{m-1}-\beta_m)}{\beta_m+1}\frac{z_m\otimes z_{m-1}^{\dagger}}{\vert z_{m-1}\vert^2}.
\label{50}
\end{equation}
On the other hand, the right-hand side of equation (\ref{tosolve}) as the explicit form
\begin{equation}
\pi_+^{\lambda,k}=(\lambda-1)\epsilon_{k+1}\frac{z_{k+1}\otimes z_k^{\dagger}}{\vert z_k\vert^2}+(\lambda-1)\epsilon_k\frac{z_k\otimes z_{k-1}^{\dagger}}{\vert z_{k-1}\vert^2}.
\label{51}
\end{equation}
Using the orthogonality of the vectors $z_k$ and comparing (\ref{50}) with (\ref{51}), we may deduce explicit expressions for the functions $\beta_k$. Indeed, we get the recurrence relations
\begin{equation}
\beta_k=\lambda-1,\quad \beta_{k-1}=(1+\beta_k)(\lambda-1)+\beta_k,\quad \beta_{m-1}=\beta_m,
\end{equation}
for $m=1,2,\cdots,k-1$. These recurrence relations are direct to solve and one gets the solution $\beta_k=\lambda-1$ and $\beta_m=\lambda^2-1$ for $m=0,1,\cdots,k-1$. We thus get an expression for $\mathcal{F}_{\lambda,k}$, solution of equation (\ref{tosolve}), given as
\begin{equation}
\mathcal{F}_{\lambda,k}=Q_k\left(\mathbb{I}-\sum_{j=0}^{k}\mathbb{P}_j+\lambda\mathbb{P}_k+\lambda^2\sum_{j=0}^{k-1}\mathbb{P}_j\right)
\end{equation}
and, for $\mathcal{F}_{\lambda,k}$ to be an element of $U(N)$, it must satisfy
\begin{equation}
\mathbb{I}=\mathcal{F}_{\lambda,k}^{\dagger}\mathcal{F}_{\lambda,k}=\mathbb{I}+(\vert\lambda\vert^4-1)\sum_{j=0}^{k-1}\mathbb{P}_j+(\vert\lambda\vert^2-1)\mathbb{P}_k\quad \iff\quad \lambda\in S^1.
\end{equation}
The second equation of (\ref{speceq}) is obtained from the first by complex conjugation and using the fact that $\mathcal{F}_{\lambda,k}\in U(N)$.

We may go further with the study of the map $\mathcal{F}_{\lambda,k}$ and relate it to a linear system of first order equations. Indeed, inspired by the Sym-Tafel formula \cite{goldstein,grundland1}, we have the property that
\begin{equation}
\mathbf{Y}_k\equiv\mathbb{P}_k+2\sum_{j=0}^{k-1}\mathbb{P}_j=\lambda\mathcal{F}_{\lambda,k}^{-1}\partial_{\lambda}\mathcal{F}_{\lambda,k}
\end{equation}
and this expression may be related to the system of linear equations for a map $\mathbf{X}_k\in su(N)$ defined as
\begin{equation}
\check{\partial}_+\mathbf{X}_k=\frac{i}{2}\pi_+^{-1,k}=-i[\check{\partial}_+\mathbb{P}_k,\mathbb{P}_k],\quad \check{\partial}_-\mathbf{X}_k=\frac{i}{2}(\pi_+^{-1,k})^{\dagger}=i[\check{\partial}_-\mathbb{P}_k,\mathbb{P}_k].
\end{equation}
This set of equations is compatible in the sense that $\{\check{\partial}_+,\check{\partial}_-\}\mathbf{X}_k=0$ if and only if $\mathbb{P}_k$ is a solution of the Euler-Lagrange equations (\ref{ELP}). Furthermore, direct calculations show that
\begin{equation}
\pi_+^{-1,k}=-2\check{\partial}_+\mathbf{Y}_k\quad \Longrightarrow\quad \mathbf{X}_k=i\left(\frac{(1+2k)}{N}\mathbb{I}-\mathbf{Y}_k\right).
\end{equation} 
In the bosonic model, the study of the maps (or $su(N)$-valued surfaces) $\mathbf{X}$ as been addressed in numerous papers \cite{delisle2,delisle3,delisle4,grundland,bolton,jiao,peng,peng1,goldstein,grundland1}. Indeed, some of them, characterized such surfaces of constant Gaussian curvature and gave a classification of solutions $\mathbb{P}$ leading to such surfaces. We have, so far, not investigated these surfaces in the supersymmetric context, but this is part of our future projects. As a preliminary observation and possible generalisation of the components of the metric $g$, we define the first fundamental form $\mathcal{I}$ of these surfaces as
\begin{equation}
\mathcal{I}=\langle d\mathbf{X},d\mathbf{X}\rangle=g_{ij}dx_idx_j+g_{i\theta_{\alpha}}dx_id\theta_{\alpha}+g_{\theta_{\alpha}i}d\theta_{\alpha}dx_i+g_{\theta_{\alpha}\theta_{\beta}}d\theta_{\alpha}d\theta_{\beta},
\end{equation}
where $i,j,\alpha,\beta=+,-$ and $g_{\mu\nu}$ are the components of the supermetric \cite{dewitt}. The inner product $\langle\cdot,\cdot\rangle$ is the usual one on $su(N)$ namely $\langle A,B\rangle=-\frac12\hbox{Tr}(AB)$ for $A,B\in su(N)$. Explicit calculations lead to
\begin{equation}
d\mathbf{X}=idx_+\check{\partial}_+\varphi_+-idx_-\check{\partial}_-\varphi_+^{\dagger}+d\theta_+(\theta_+\check{\partial}_+\varphi_++i\varphi_+)-d\theta_-(\theta_-\check{\partial}_-\varphi_+^{\dagger}+i\varphi_+^{\dagger}),
\end{equation}
where $\varphi_+$ was previously defined in (\ref{conserve}) and this allows us to give explicit expressions for the supermetric components $g_{\mu\nu}$:
\begin{eqnarray}
g_{++}&=&\langle m_+,m_+\rangle,\quad g_{--}=\langle m_+^{\dagger},m_+^{\dagger}\rangle,\quad g_{+-}=\langle m_+,m_+^{\dagger}\rangle,\quad g_{\theta_+\theta_-}=\langle\rho_+^{\dagger},\rho_+\rangle\nonumber\\
g_{+\theta_+}&=&\langle m_+,\rho_+\rangle,\quad g_{+\theta_-}=\langle m_+,\rho_+^{\dagger}\rangle,\quad g_{-\theta_+}=-\langle m_+^{\dagger},\rho_+\rangle,\quad g_{-\theta_-}=\langle m_+^{\dagger},\rho_+^{\dagger}\rangle,
\end{eqnarray}
where $m_+=i\check{\partial}_+\varphi_+$ and $\rho_+=\theta_+\check{\partial}_+\varphi_++i\varphi_+$. The other components of the supermetric are deduced from the facts that $g_{ij}=g_{ji}$, $g_{i\theta_{\alpha}}=g_{\theta_{\alpha}i}$ and $g_{\theta_{\alpha}\theta_{\beta}}=-g_{\theta_{\beta}\theta_{\alpha}}$. We can go further in the analysis of these components. We have
\begin{equation}
\rho_+=\check{\partial}_+(\varphi_+\theta_+)
\end{equation}
and, using the Taylor expansion $\varphi_+=\xi_0+i\theta_+A_++i\theta_-A_--\theta_+\theta_-\xi_{+-}$, we deduce an explicit expression for $\rho_+$ given as
\begin{equation}
\rho_+=i\xi_0-\theta_-A_-.
\end{equation}
In this formulation, the fields $A_{\mu}=A_{\mu}(x_+,x_-)$ are bosonic and $\xi_{\mu}=\xi_{\mu}(x_+,x_-)$ are fermionic. The components $\xi_0$ and $A_-$ may be explicitly calculated. Indeed, we have
\begin{equation}
\xi_0=\varphi_+\vert_{\theta_+=\theta_-=0},\quad A_-=\check{\partial}_-\varphi_+\vert_{\theta_+=\theta_-=0}.
\end{equation}
Using the Taylor expansion for the orthogonal set of vectors $\{z_j\}$ as
\begin{equation}
z_j=u_j+i\theta_+\mu_j+i\theta_-\nu_j-\theta_+\theta_-v_j, \quad j=0,1,\cdots,N-1,
\end{equation}
we deduce the orthogonality of the set of vectors $\{u_j\}$ and, using the notation $\varphi_+^j=i[\mathbb{P}_j,\check{\partial}_+\mathbb{P}_j]$, we get
\begin{eqnarray}
\xi_0^j&=&-i\left(\epsilon_{j+1}^f\frac{u_{j+1}\otimes u_j^{\dagger}}{\vert u_j\vert^2}+\epsilon_j^f\frac{u_j\otimes u_{j-1}^{\dagger}}{\vert u_{j-1}\vert^2}\right),\\
A_-^j&=&i\vert\epsilon_{j+1}^f\vert^2\frac{\vert u_{j+1}\vert^2}{\vert u_j\vert^2}\left(\check{\mathbb{P}}_j-\check{\mathbb{P}}_{j+1}\right)+i\vert \epsilon_j^f\vert^2\frac{\vert u_j\vert^2}{\vert u_{j-1}\vert^2}\left(\check{\mathbb{P}}_{j-1}-\check{\mathbb{P}}_j\right),
\end{eqnarray}
where $\epsilon_k=\epsilon_k^f+i\theta_+\epsilon_k^b$ and $\check{\mathbb{P}}_k=\frac{u_k\otimes u_k^{\dagger}}{\vert u_k\vert^2}$ are hermitian orthogonal projectors of rank one for all $k$. This allows us to calculate the explicit expression of the component $g_{\theta_+\theta_-}^j$ and we get
\begin{equation}
g_{\theta_+\theta_-}^j=-\frac12\left(\vert\epsilon_{j+1}^f\vert^2\frac{\vert u_{j+1}\vert^2}{\vert u_j\vert^2}+\vert \epsilon_j^f\vert^2\frac{\vert u_j\vert^2}{\vert u_{j-1}\vert^2}\right)+\theta_+\theta_-\vert\epsilon_j^f\vert^2\vert \epsilon_{j+1}^f\vert^2\frac{\vert u_{j+1}\vert^2}{\vert u_{j-1}\vert^2}.
\end{equation}
In the hypothesis, where the above quantities are element of a one-dimensional complex Grassmann algebra \cite{dewitt,cornwell} spanned by the generator $\eta_+$ with $\eta_+^{\dagger}=\eta_-$ and $\eta_+=\eta_1+i\eta_2$, we can write $\epsilon_k^f=a_k^+\eta_++a_k^-\eta_-$, $u_k=b_k+\eta_+\eta_-b_k^{+-}$ and $\theta_+=c_+\eta_++c_-\eta_-$ for $a_k^{\mu}$ and $c_{\mu}$ commuting scalar quantities and $b_k^{\mu}$ bosonic vectors. In this special case, the supermetric component $g_{\theta_+\theta_-}^j$ take the explicit form
\begin{equation}
g_{\theta_+\theta_-}^j=\frac{\eta_+\eta_-}{2}\left((\vert a_{j+1}^+\vert^2-\vert a_{j+1}^-\vert^2)\frac{\vert b_{j+1}\vert^2}{\vert b_j\vert^2}+(\vert a_j^+\vert^2-\vert a_j^-\vert^2)\frac{\vert b_j\vert^2}{\vert b_{j-1}\vert^2}\right).
\end{equation}
This metric component was studied in \cite{bertrand} to describe conformal maps in the supersymmetric context.

\section{Future outlooks}

In this paper, we have presented a new perspective of the two-dimensional supersymmetric $\mathbb{C}P^{N-1}$ sigma model from a supergeometric \cite{dewitt,cornwell} point of view. Indeed, from a theorem of Khemar \cite{khemar} regarding superharmonic maps, we have classified the $1$-superforms which are of the Maurer-Cartan form and related to the $\mathbb{C}P^{N-1}$ model. From these Maurer-Cartan $1$-superforms, we have extended our analysis to the associated linear spectral problem and found its explicit solutions in terms of the recently discovered solutions of the supersymmetric $\mathbb{C}P^{N-1}$ model. The solutions of the linear spectral problem where used, from the Sym-Tafel \cite{goldstein,grundland1} and Weierstrass immersion \cite{kono} formulas, to determine $su(N)$-valued maps (or surfaces) which are solutions of an integrable system of partial differential equations.

In the bosonic model, the geometry of the $su(N)$-valued surfaces where extensively studied. In the supersymmetric context, we have constructed for the first time this kind of surfaces and a complete study of the geometry of these surfaces remains. This perspective is part of our future outlooks such as the study of constant Gaussian curvature surfaces. In this paper and in \cite{bertrand}, preliminary observations and generalisations were made on this subject. 

As a final comment, in \cite{delisle}, new solutions were presented for this model but the density of this set was not shown. So, we may have missed some solutions. Showing the density of this set of solutions, described in section 2, is an open problem and remains part of our current research.
\section*{Acknowledgement}
The author acknowledges a Natural Sciences and Engineering Research Council of Canada (NSERC) postdoctoral fellowship. The author would like to personally thank V\'eronique Hussin, Wojtek J. Zakrzewski and Fr\'ed\'eric H\'elein for helpful discussions and remarks.

\end{document}